# Effect of non-magnetic impurities on the gap of a $d_{x^2-y^2}$ superconductor as seen by angle-resolved photoemission

R. Fehrenbacher

*Max-Planck-Institut für Festkörperforschung, Heisenbergstr. 1, D-70569 Stuttgart, Germany*

(March 15, 1996)

An analysis of angle-resolved photoemission (ARPES) experiments in the superconducting state of the high $T_c$ copper-oxides is presented. It is based on a phenomenological weak-coupling BCS model which incorporates the experimental normal state dispersion extracted from ARPES, and non-magnetic impurity scattering in the presence of a $d_{x^2-y^2}$ order parameter (OP). It is shown that already in the pure case, the broadening by finite momentum resolution of the analyzer leads to a finite region of apparent 'gaplessness' around the true node of the OP. Non-magnetic impurities further amplify this effect by introducing additional spectral weight around zero frequency. At sufficiently large impurity concentrations $n_i \approx 0.02 - 0.05$, this results in an extended region of 'gaplessness' up to $\delta\phi = \pm 7$ ($\phi$ the angle on the Fermi surface) around the true node for a large range of moderate to strong impurity potential strengths. Different ways to identify the presence of impurity scattering in the ARPES spectra are proposed.

PACS numbers: 74.20.-z 74.20.Mn, 74.25.Nf, 74.62.Dh

## I. INTRODUCTION

Major experimental progress has recently been made in identifying the order parameter (OP) symmetry of the high $T_c$ cuprates. A larger number of results strongly suggest that the carriers pair with a non-trivial, spatially anisotropic $d_{x^2-y^2}$ symmetry.[1] Among these probes, angle-resolved photoemission (ARPES) has the unique potential of determining the superconducting gap as a function of momentum, and indeed, strong support for the $d$-wave scenario has recently been reported using this technique.[2,3]

Since the cuprates are the first materials in which ARPES experiments were able to detect a superconducting gap, there is not much theoretical work experimentalists can compare their spectra to, and hence in most cases the extraction of the gap is done on a rather qualitative level. It is therefore important to provide some minimal theoretical guidance, and put the data analysis on firmer grounds. To this end we calculate so-called energy distribution curves (EDC) using a rather simple phenomenological BCS model which consists of the ARPES normal state quasiparticle dispersion supplemented by a $d_{x^2-y^2}$ pair interaction and non-magnetic impurity scattering.[4] The effects of impurity scattering are of particular importance, since the reduced symmetry of the OP allows for severe pair-breaking even in the non-magnetic case.[5]

One rather selective signature of the $d$-wave state, as compared to a strongly anisotropic s-wave state for instance, is the possibility of so-called resonant scattering by non-magnetic impurities.[6,7] If the impurity potential is such that the resonant condition is satisfied, a very small impurity concentration $n_i$ is sufficient to create a large residual density of single particle states (DOS) $N_{res}$ at the Fermi level. Naturally, this effect also has to manifest itself in the spectral function $A(\mathbf{k},\omega)$, and hence it is of interest to study whether it could be observed in an ARPES experiment. Even away from resonance, the impurities induce a finite $N_{res}$, albeit the growth with the number of impurities is much smaller. Nevertheless, at sufficiently large concentrations of a few percent, one expects to see an effect on $A(\mathbf{k},\omega)$ in this limit also. In earlier work,[8] we already studied the influence of impurity scattering on ARPES spectra from a slightly different perspective, but the model used there did not incorporate a realistic normal state quasiparticle dispersion. When comparing to experimental spectra, however, a realistic model for the dispersion is crucial, since it dramatically affects resolution broadening, as we shall discuss below.

By analogy with the rise of $N_{res}$ as a function of impurity concentration, one might expect that the effect on the gap as extracted by ARPES is also to progressively wipe it out as impurities are added. Indeed, we shall show that, depending on how the gap is extracted from the ARPES spectra, there can be a finite region on the Fermi surface around the $d_{x^2-y^2}$ node, where the measured gap appears to vanish, even though the order parameter is finite. However, to a large extent, this effect is caused by the finite momentum resolution of the analyzer. It occurs even for the case of a pure BCS spectral function, although the 'gapless' region is then much smaller.

The plan of this article is as follows: In section II, we shall start with a brief discussion of the photoemission process, and discuss the influence of finite experimental resolution in both energy and momentum on the spectral line shape. Based on this, in section III, we shall discuss different methods to extract the superconducting gap from the ARPES spectra, and show how well its resulting $\mathbf{k}$ dependence compares with the true OP. In section IV, the influence of non-magnetic impurity scattering on the line shape, as well as on the extracted gap will be elucidated. Finally,

1

a comparison with experiments, and a critical discussion will be presented in section V.

## II. SPECTRAL FUNCTION INTERPRETATION OF ARPES

In the analysis of ARPES spectral line shapes, we assume the validity of the spectral function interpretation. In other words, we assume that the ARPES intensity (neglecting effects of finite experimental resolution) is given by $I(\mathbf{k},\omega) = I_0(\mathbf{k})f(\omega)A(\mathbf{k},\omega)$. The prefactor $I_0(\mathbf{k})$ depends on incident photon energy, polarization, and on the final state through the electron-photon matrix element, however, it is only weakly dependent on $\omega$ or temperature. Hence, the line shape is entirely determined by the Fermi function $f(\omega)$, and $A(\mathbf{k},\omega)$. We shall later comment on this simplifying assumption, and show that it leads to a self-consistent interpretation of the ARPES experiments on $Bi_2Sr_2CaCu_2O_8$ (Bi2212), which is the best studied cuprate. Recently, it has been demonstrated that this approach works also well in the analysis of other quantities.[9]

A slight complication regarding the interpretation of experimental spectra results from the finite resolution of the detector in both frequency and momentum. Usually, the detector probes a finite momentum window $\delta\mathbf{k}$ with constant probability distribution due to its finite extent. Apart from the physical dimension of the detector, the size of this window depends also on the incident photon energy. A circular window of radius $0.036\pi/a$, $a$ the planar lattice constant, is a typical dimension, realistic for the measurements from Ref. 3 performed at incident photon energy 22eV. All calculations in this paper were done using this $\mathbf{k}$ resolution, and the temperature entering the Fermi function was fixed at $T = 10K$, a typical value for actual measurements in the superconducting state. The frequency resolution of the detector is a Gaussian, and current state-of-the-art devices can reach a standard deviation $\sigma_\omega$ as low as 7meV. Taking both resolution effects into account, the ARPES intensity or EDCs should be given by[3]

$$I(\mathbf{k},\omega) = I_0 \int_{\delta\mathbf{k}} d\mathbf{k}' \int d\omega' R(\omega - \omega') f(\omega') A(\mathbf{k}',\omega'), \quad (1)$$

where $R(\omega - \omega') \sim \exp\left(-\frac{(\omega-\omega')^2}{2\sigma_\omega}\right)$. Here we neglected the $\mathbf{k}$ dependence of $I_0$.

It is very instructive to calculate EDCs according to this formula within simple BCS theory. Then the spectral function takes the familiar form

$$A(\mathbf{k},\omega) = u_\mathbf{k}^2 \delta(\omega - E_\mathbf{k}) + v_\mathbf{k}^2 \delta(\omega + E_\mathbf{k}), \quad (2)$$

with $E_\mathbf{k} = \sqrt{\xi_\mathbf{k}^2 + \Delta_\mathbf{k}^2}$ the quasiparticle energy, and $u_\mathbf{k}^2 = (1 + \xi_\mathbf{k}/E_\mathbf{k})/2$, $v_\mathbf{k}^2 = (1 - \xi_\mathbf{k}/E_\mathbf{k})/2$ the coherence factors. To be as realistic as possible, for the quasiparticle dispersion $\xi_\mathbf{k}$, we use a tight-binding fit to normal state ARPES data on Bi2212[3,10] with real space hopping matrix elements $[t_0,\cdots,t_5] = [0.879, -1, 0.28, -0.087, 0.094, 0.087]$, ($t_0$ on-site, $t_1$ nn, $t_2$ nnn hopping, ... ).[11,12] All energies are measured in units of $|t_1| = 0.149$eV. The value of $t_0$ corresponds to a hole doping of $\delta = 0.17$. As suggested by the ARPES data, the fitted quasiparticle band-structure exhibits a saddle point at $\mathbf{k} = (\pi,0)$ with the resulting van Hove singularity located at approximately 30meV below $\varepsilon_F$ in Bi2212. Normal state data on other hole-doped cuprates show a similar dispersion.[13]

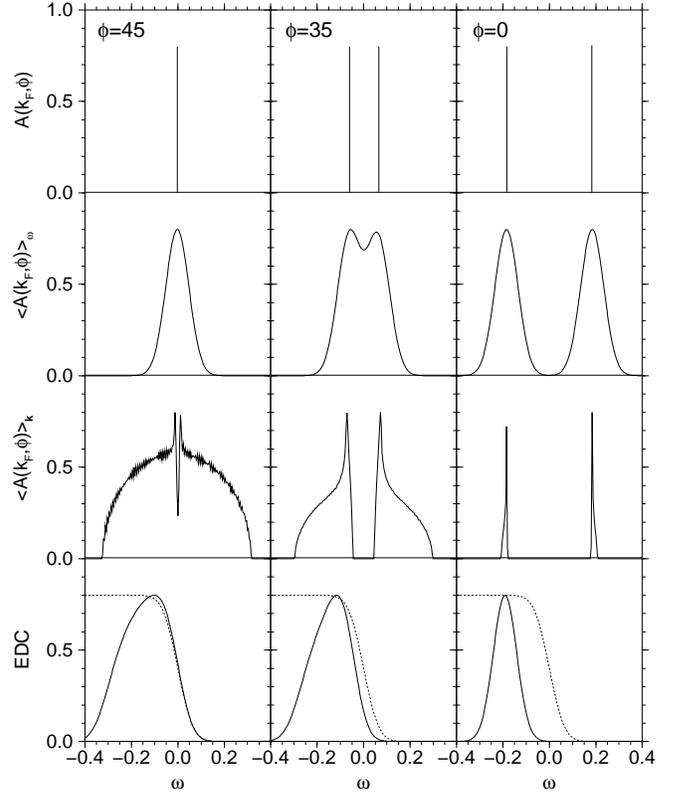

**Fig. 1.** The effect of the finite experimental resolution on the EDCs as calculated from the pure BCS $d_{x^2-y^2}$ spectral function. The three columns are calculations for different angles on the Fermi surface, $\phi = 45, 35, 0$ from left to right. Top row: the pure spectral function. Second row: spectral function after frequency broadening. Third row: spectral function after momentum broadening. Last row: EDC calculated from (1), and normalized to the spectrum of a resolution broadened Fermi function (dotted curve).

The OP is chosen as $\Delta_\mathbf{k} = \Delta(\cos k_x - \cos k_y)/2$, and the size of the maximum gap is fixed at $\Delta = 0.2$ (for $T = 0$), which corresponds to approximately 30meV, consistent with ARPES results on Bi2212.[2,3] Fig. 1 demonstrates the effect of the experimental resolution on the ARPES line shape. We show spectra at three different angles $\phi = 0, 35, 45$ on the Fermi surface. $\phi$ is measured relative to the $(\pi,\pi)$ point, around which the Fermi surface is closed. The top row displays the pure BCS spectral function, i.e., two delta functions provided the OP is finite (the node in the $d_{x^2-y^2}$ gap is at $\phi = 45$). In the second row, the effect of the frequency resolution is shown, which simply leads to a Gaussian broadening of the peaks. Of course this broadening does not depend on the Fermi surface angle. This is very different though for the broadening caused by the finite **k** resolution, as shown in row three. The shape of the spectra is similar to a BCS DOS for an isotropic gap, which is not surprising, since the order parameter does not vary too much within the resolution **k** window over which one integrates. The broadening is clearly strongly dependent on **k**. It is largest at $\phi = 45$ (the region where the $d_{x^2-y^2}$ gap has a node), and hardly noticeable at $\phi = 0$ where the $d_{x^2-y^2}$ gap is maximal. This effect is a simple consequence of the non-trivial dispersion: Around $\phi = 45$, the quasiparticle band is very dispersive, hence the finite momentum window of the analyzer samples **k** points with a wide spread in energy. This results in a large broadening. The situation is reversed around $\phi = 0$ ($[\pi, 0]$), where due the saddle point, the band is only weakly dispersive, and hence the **k** points in the resolution window all have similar energies leading to extremely small broadening. Note, that the **k** dependence of the OP itself further amplifies this difference in momentum broadening, since it has a linear dependence on **k** around $\phi = 45$ (large variation of $\Delta_\mathbf{k}$ within $\delta \mathbf{k}$), and a quadratic one near $\phi = 0$ (small variation of $\Delta_\mathbf{k}$ within $\delta \mathbf{k}$).

Looking at the EDCs calculated according to (1), and displayed in the bottom row of Fig. 1, one clearly observes that the width of the peaks is about twice as large at $\phi = 45$ as compared to $\phi = 0$ due the momentum resolution effect. From this, it is also evident that the widths are **k** resolution limited around $\phi = 45$, whereas at $\phi = 0$ they are frequency resolution limited. The absolute values of the peak width are in good agreement with the experimental spectra of Ref. 3. This justifies a posteriori the assumption of interpreting the ARPES data using a simple spectral function approach, and also shows that the values claimed by the experimentalists about their resolution are self-consistent.

## III. SUPERCONDUCTING GAP

Next, we are interested in the values of the superconducting gap which is obtained by extraction from the calculated EDCs. In the experimental literature three different methods have been employed: (i) Measuring the shift of the leading edge of the EDC in the superconducting state as compared to the normal state EDC at the same **k** point. This method is not really suitable to obtain a quantitative measure of the gap since the two compared spectra are neither taken at the same temperature, nor time. Furthermore, the normal state EDCs are substantially broader than the superconducting ones leading to an additional uncertainty. Therefore, this method can only give an indication that a superconducting gap has opened up, but is not trustable for an estimate of its size.

(ii) Measuring the shift of the leading edge of the EDC in the superconducting state as compared to an angle-averaged spectrum from a non-superconducting reference sample such as Ag or Pt which is in electrical contact with the superconductor. The spectra of the reference sample essentially represent the occupied density of single particle states, and should therefore be well represented by a resolution broadened Fermi function at the temperature of the experiment. This method allows for a quantitative estimate of the gap, and according to our calculations should always give a lower bound of its true value. It is not biased, *i.e.*, no assumption about the form of the spectral function is necessary. However, we shall show that the error as compared to the true OP can depend quite strongly on the shape of $A(\mathbf{k}, \omega)$.

(iii) Finally, one can try a fit of the experimental EDC to a theoretical curve based on Eq. (1). However, this requires the knowledge of $A(\mathbf{k}, \omega)$, which is usually assumed to be of the simple BCS form. Hence it can only be reliable if lifetime effects which could alter the BCS $A(\mathbf{k}, \omega)$ are negligible. Furthermore, one has to make an assumption about the frequency dependence of the background in the spectra which leads to additional uncertainty, even though one can check the consistency of the assumptions. Nevertheless, this is by far the most accurate way to determine the gap.

In this article, we shall use method (ii) and extract the shift $\Delta_{sh}$ of our calculated EDCs compared to a resolution broadened Fermi function given by $N_R(\omega, T) = \int d\omega' R(\omega - \omega') f(\omega', T)$ which is representative for the spectrum of the reference sample. The shift is evaluated at half the peak value after normalizing the EDC and $N_R(\omega, T)$ such that the maximum of the EDC coincides with the constant value of $N_R(\omega, T)$ at large negative $\omega$. This defines the meaning of a normalized spectrum for this article. In the bottom panel of Fig. 1, we have illustrated this procedure. Note, that due to the finite momentum resolution it is actually possible to obtain negative shifts by this procedure, especially if the resolution **k** window is rather small. This is a consequence of the non-trivial $\omega$ dependence resulting from the **k** averaging as shown in row three of Fig. 1.

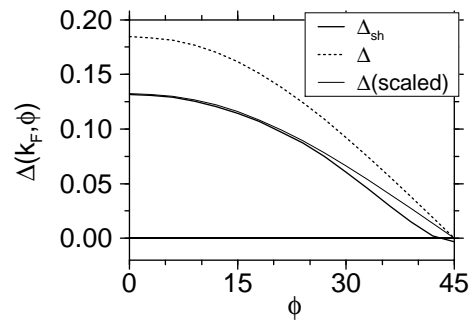

**Fig. 2.** The superconducting gap on the Fermi surface as a function of the angle $\phi$ for the clean case. The true OP $\Delta$ is compared with $\Delta_{sh}$, which is the shift of the leading edge of the EDC compared to the reference spectrum $N_R(\omega, T)$.

In Fig. 2, we compare the extracted shift with the true OP as a function of angle $\phi$ on the Fermi surface. In the large gap region, the $\phi$ dependence of $\Delta_{sh}$ pretty much follows that of the true OP, however, around the node, there is a big discrepancy coming from the finite **k** resolution: Whereas the $d_{x^2-y^2}$ OP goes to zero linearly, $\Delta_{sh}$ takes a super-linear dependence, and becomes zero already at an angle of $\phi \approx 42$, producing an apparent extended region of nodes around $\phi = 45$. Also, the gap is underestimated by a factor of at least 1.4 even in the large gap region which is of course obvious, since the true gap is given not by the shift but by the position of the peak in the EDC. This example illustrates that the **k** dependence of the gap around the node region is very difficult to quantify using the shift method already in a simple BCS model. In the following section, we shall show that impurity effects can even further blur the true value of the gap through lifetime effects.

## IV. IMPURITY EFFECTS

As mentioned in the introduction, non-magnetic impurities have a severe pair-breaking effect in a *d* wave superconductor. In this section, we shall investigate their effect on the spectral function, the EDC, and the values $\Delta_{sh}$. We use a short-range potential $V_i(\mathbf{r}) = u\delta(\mathbf{r} - \mathbf{r}_i)$, and apply the self-consistent t-matrix approximation.[6,7] This approach has recently been shown to yield accurate results in the dilute impurity limit.[14] Working in particle-hole space, the t-matrix $\widehat{T}$ satisfies the following Lippmann-Schwinger equation (quantities with a hat represent matrices)

$$\widehat{T}(\omega) = u \left[ \widehat{\sigma}_3 + \widehat{\sigma}_3 \widehat{T}(\omega) \frac{1}{\Lambda} \sum_\mathbf{k} \widehat{g}(\mathbf{k},\omega) \right]. \quad (3)$$

Here we introduced the propagator

$$\widehat{g}(\mathbf{k},\omega) = \frac{\widetilde{\omega}\widehat{\sigma}_0 + \widetilde{\Delta}_\mathbf{k}\widehat{\sigma}_1 + \widetilde{\xi}_\mathbf{k}\widehat{\sigma}_3}{\widetilde{\omega}^2 - \widetilde{\Delta}_\mathbf{k}^2 - \widetilde{\xi}_\mathbf{k}^2}, \quad (4)$$

where $\widetilde{\omega} = \omega - \Sigma_0, \widetilde{\Delta}_\mathbf{k} = \Delta_\mathbf{k} + \Sigma_1, \widetilde{\xi}_\mathbf{k} = \xi_\mathbf{k} + \Sigma_3, \xi_\mathbf{k}$ the quasiparticle energy. The self-energy (like all other matrices) is expanded in terms of the identity and Pauli matrices $\widehat{\sigma}_0 \cdots \widehat{\sigma}_3$ as $\widehat{\Sigma} = \Sigma_j \widehat{\sigma}_j$, and is given by $\widehat{\Sigma} = n_i \widehat{T}$, $n_i$ the impurity concentration, and $\Lambda$ the volume. For the $d_{x^2-y^2}$ OP, the off-diagonal self-energy vanishes by symmetry, and from (3), the non-zero components are

$$\Sigma_0 = \frac{n_i G_0}{(1/u - G_3)^2 - G_0^2}; \quad \Sigma_3 = \frac{n_i(1/u - G_3)}{(1/u - G_3)^2 - G_0^2}. \quad (5)$$

where $G_i(\omega) = \Lambda^{-1} \sum_\mathbf{k} g_i(\mathbf{k},\omega)$. As outlined previously,[4] we take into account the full structure of the t-matrix, *i.e.*, we do not assume particle-hole symmetry. The self-consistency equations (5) are solved numerically on a $d \times d$ **k**-space grid with $d \leq 4000$. The spectral function is then evaluated from

$$A(\mathbf{k},\omega) = -\frac{\text{sgn}\omega}{\pi} \text{Im}\left\{ \widehat{g}_{11}(\mathbf{k},\omega) \right\}, \quad (6)$$

and the EDC is obtained using Eq. (1).

As shown earlier,[4] the effect of the non-trivial DOS is rather dramatic, in particular if a vHs is close to $\varepsilon_F$, as in the present case: (i) The s-wave scattering phase shift $\delta_0$ acquires a strong frequency dependence already in the *normal* state, which also reflects itself in the scattering rate, hence resistivity. Furthermore, the dependence on *u* is highly non-trivial. (ii) In the self-energy Eqs. (5), the cotangent $c = \cot\delta_0$ (being *frequency independent* in case of a constant normal DOS) that is usually used to parameterize the scattering strength, is replaced by the (now *frequency dependent*) quantity $\widetilde{c}(\omega) \equiv -1/u + G_3(\omega)$. This leads to a strong sensitivity of the superconducting DOS and $T_c$ on both *u* and the chemical potential. Resonant scattering is usually observed for $|c| \ll 1$ (corresponding to $|u| \approx \infty$ for constant DOS). In the present case, this translates to the condition $\widetilde{c}(\omega = 0) \ll 1$, allowing for resonant scattering even if $|u| \ll \infty$.

While the case of resonant scattering is of particular importance for quantities which are very sensitive to $N_{\text{res}}$, such as the temperature dependence of the NMR relaxation rate, or the London penetration depth, for ARPES experiments it is not so crucial. The difference between resonant and non-resonant scattering is most dramatic at small impurity concentrations. However, then the self-energy is also not so large, and the finite resolution in the ARPES experiments smears out the sharp feature at resonance which exists in the unaveraged spectral function. In order to see a clear effect in the EDC, it is necessary to go to rather large impurity concentrations $n_i \geq 0.02$. Hence, in this article, we shall concentrate on this case, and consider moderate to strong impurity potential strengths *u*.

Fig. 3 shows a comparison of the unaveraged spectral function on the Fermi surface for $n_i = 0.05$ between the normal state and the superconducting state at $\phi = 0, 45$ for various potential strengths. The normal state Fermi surface was determined by the criterion that $n(k_F) = \int_{-\infty}^0 d\omega A(k_F,\omega) = 1/2$, *i.e.*, we took the shift of the chemical potential caused by the impurity scattering into account (of course this shift is only small, of the order of $t_1/50$ for $n_i = 0.05$, and dependent on the magnitude *and* sign of *u*). This criterion ensures that the peak of the normal state spectral function is at $\omega = 0$. Concentrating first on the spectra for $\phi = 45$ (at the node), one notes that there is a strong dependence of $A(\mathbf{k},\omega)$ on *u*. For most values there is a clear double peak structure, one peak above, one peak below $\varepsilon_F$. The relative strength of the peaks varies with *u*. As $1/u$ is decreased from positive to negative values, the peak above $\varepsilon_F$ loses, and the one below $\varepsilon_F$ gains spectral weight. At $1/u \approx 0.0$ the two peaks have equal strength, and the spectra looks like a single very broad peak.

This behavior can be readily understood by examining Fig. 4 where we plot $\text{Im}\{\widehat{\Sigma}_{11}(\omega)\}$, which is responsible for the quasiparticle damping. We observe a single peak structure on a slowly varying background, which moves from negative to positive frequencies as $1/u$ is varied from 0.4 to -0.4. This structure leads to the small peak in $A(k_F, \omega)$ provided it is away from $\varepsilon_F$. When it is at $\varepsilon_F$, the case which corresponds to resonant scattering, the self-energy peak coincides with the quasiparticle peak positioned at $\xi_\mathbf{k}$ leading to a single structure in $A(k_F, \omega)$.

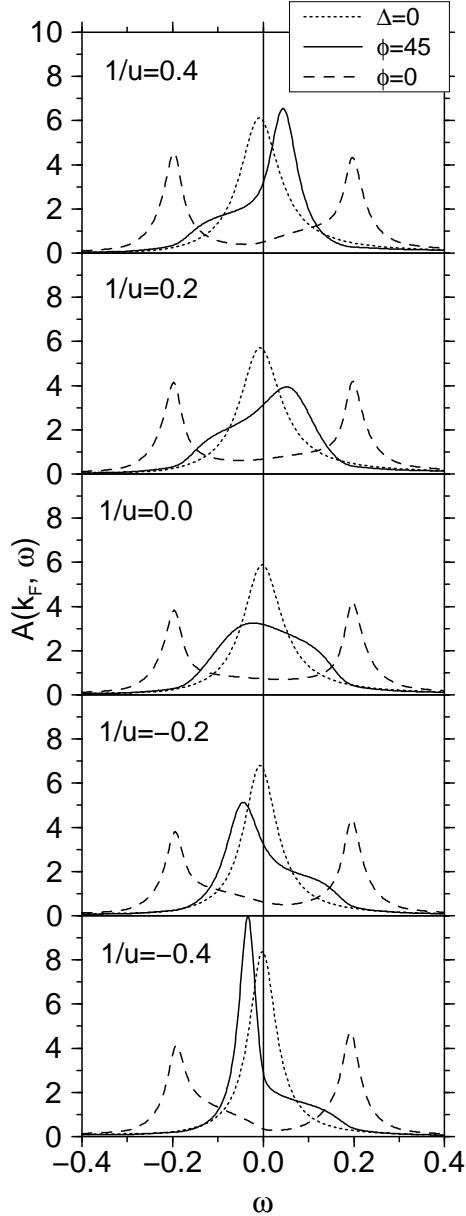

**Fig. 3.** Dependence of $A(k_F, \phi)$ on the strength of the impurity potential $u$ for $n_i = 0.05$. Spectra are shown in the normal as well as in the superconducting state (at the node $\phi = 45$, and the maximum gap $\phi = 0$).

At $\phi = 0$ (the large gap region) the spectra consists of the two peaks positioned at $\pm \Delta_\mathbf{k}$ as expected from the BCS form of $A(\mathbf{k}, \omega)$. The impurity scattering substantially broadens the peaks, and it also leads to spectral weight at zero frequency, which is responsible for the finite $N_{res}$. The weight around $\omega = 0$ depends again on the value of $1/u$, for positive (negative) $1/u$ it is concentrated at positive (negative) $\omega$.

One further point worth mentioning is the fact that the impurity scattering violates the approximate sum rule stating that $n(k_F)$ should be independent of temperature above and below $T_c$.[9] It is evident from Fig. 3 that compared to its value in the normal state (= 0.5 by definition), $n(k_F)$ in the superconducting can grow or decrease depending on the value of $1/u$. The effect is particularly severe at $\phi = 45$. Here, of course this is due to the strong particle-hole asymmetry introduced by the impurity scattering and the normal state dispersion, under which condition the sum rule is not applicable.[9] It might be that this effect could shed some light on the puzzling results from Ref. 15, where a large change in the ARPES spectral weight was observed upon cooling through $T_c$.

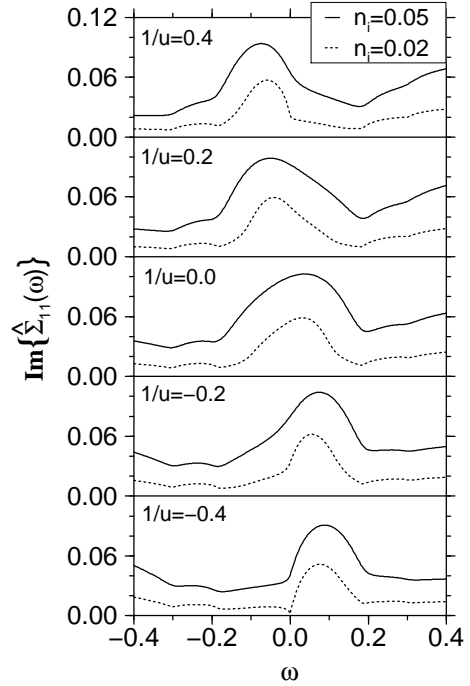

**Fig. 4.** Dependence of the imaginary part of the self-energy component $\text{Im}\{\hat{\Sigma}_{11}\}$ on the strength of the impurity potential $u$ for $n_i = 0.02, 0.05$.

Next, we look at the effect of the impurity scattering on the EDCs as calculated using (1). Fig. 5 illustrates the influence of the scattering on the shift $\Delta_{sh}$, at $1/u = -0.4$ for four angles $\phi = 0, 35, 40, 45$ on the Fermi surface, and at two values of $n_i = 0.02, 0.05$. For comparison, the raw spectral functions are also shown in column one. At the node, $\phi = 45$ the impurities have virtually no influence on the EDC as evident by comparing with the spectrum in Fig. 1. However, at $\phi = 40$, increasing the impurity concentration leads to a progressive closing of the gap measured in terms of the shift: at $n_i = 0.05$, $\Delta_{sh} = 0$. The same effect is seen further away from the node at $\phi = 35$. Here, $\Delta_{sh}$ is still finite at $n_i = 0.05$, but is substantially reduced from its value at $n_i = 0.02$.

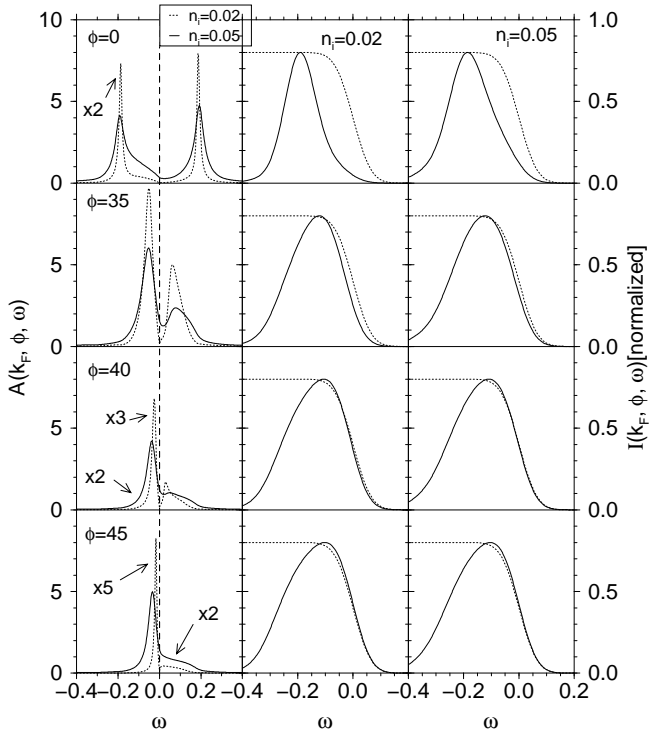

**Fig. 5.** Spectral function, and the corresponding EDC calculated using (1) for $1/u = -0.4$ and $n_i = 0.02, 0.05$ at four different angles on the Fermi surface, $\phi = 0, 35, 40, 45$ (from top to bottom). First column: the unaveraged spectral functions. Second column: EDCs for $n_i = 0.02$. Third column: EDCs for $n_i = 0.02$. The dotted line in columns two and three represents the resolution broadened Fermi function $N_R(\omega, T)$.

At $\phi = 0$, where the OP reaches its maximum, the shift is also strongly reduced due the progressive broadening of the EDC by the impurity scattering, but at $n_i = 0.05$, a new effect appears: the leading edge slope of the EDC does not match the slope of the resolution function $N_R(\omega, T)$. This behavior is a consequence of the enhanced spectral weight around zero frequency which also induces the large $N_{res}$. The slope problem is most critical for scattering strengths around the resonance limit.

Fig. 6 addresses this issue in more detail. Here, we plot the EDCs at $\phi = 0$, $n_i = 0.0, 0.05$ for various values of $1/u$. For rather weak scattering ($1/u = 1.0, -1.0$), the slopes of the EDCs and the resolution function are pretty much parallel. However, in the strong scattering limit, mostly pronounced at $1/u = 0.0$, a low frequency tail in the EDC makes the peak substantially asymmetric, and results in the above mentioned slope problem. This effect could be a way to identify strong impurity scattering in ARPES experiments. We emphasize however, that it would be seen most easily in regions where the OP is large, i.e., around $(\pi, 0)$.

In fact, a similar behavior has recently been reported from ARPES measurements on electron irradiated samples of Bi2212.[16] The electron irradiation is believed to create Frenkel-type point defects by displacing oxygen atoms in the $CuO_2$ plane. This type of defect apparently does not lead to a Curie tail in the magnetic susceptibility,[17] hence it seems to represent a non-magnetic potential scattering center. The spectrum shown in Ref. 16 (measured in the large gap region) indeed seems to have acquired a low-frequency tail as expected from the theory. A more detailed study would be desirable to unambiguously identify the slope problem.

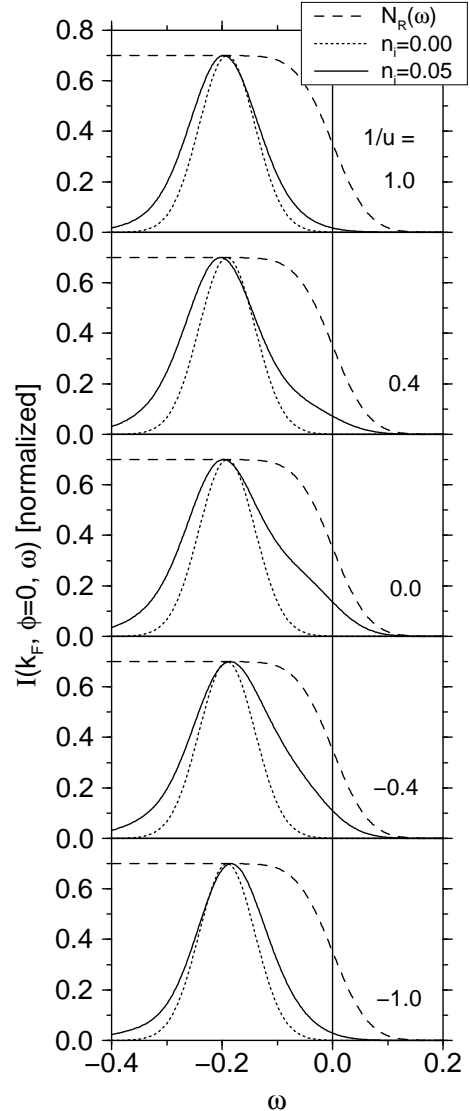

**Fig. 6.** EDCs for various values of $1/u$ and $n_i = 0.05$ at the gap maximum $\phi = 0$, compared to the pure spectra, and the resolution broadened Fermi function $N_R(\omega, T)$. The slope problem is obvious close to resonance ($1/u = 0.0, -0.4$).

Fig. 6 also clearly illustrates the problems one encounters when trying to extract the OP from the leading edge shift $\Delta_{sh}$. Even though the OP is precisely the same for all spectra in Fig. 6, the value one extracts from $\Delta_{sh}$ have a large variation. Such a slope problem will always occur if the underlying spectral function deviates substantially from a single-peak structure.

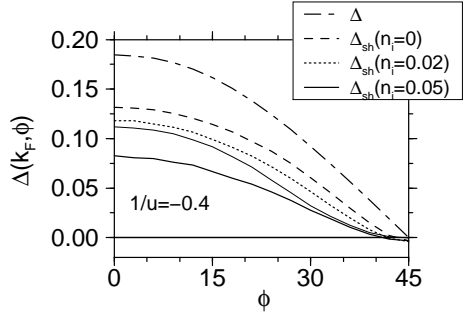

**Fig. 7.** The superconducting gap on the Fermi surface as a function of the angle $\phi$. The true OP $\Delta$ is compared with $\Delta_{sh}$, calculated for $1/u = -0.4$ and different impurity concentrations $n_i = 0.0, 0.02, 0.05$. The thin solid line shows $\Delta_{sh}$ evaluated for $1/u = 0.4, n_i = 0.05$.

In Fig. 7, we show the angle dependence of $\Delta_{sh}$ on the Fermi surface for $1/u = -0.4$. As compared to the true OP, the shift $\Delta_{sh}$ extracted from the EDC clearly underestimates the value of $\Delta$ already in the pure case, an effect which is strongly enhanced by the broadening caused by impurity scattering. Furthermore, the impurities lead to a 'gapless' region around $\phi = 45$ of size up to $\delta\phi = \pm 7$, where the shift $\Delta_{sh}$ is essentially zero. Such behavior has been reported earlier for some samples measured by ARPES, while it is not seen in others.[18] Our results strongly suggest that it is the sample purity which is responsible for these differences. A non-linear dependence on $\phi$ of the OP itself is not necessary to explain the flatness around $\phi = 45$.

The thin line in Fig. 7, was obtained for $1/u = 0.4, n_i = 0.05$, and shows (in comparison to the curve for $1/u = -0.4$) that the shift $\Delta_{sh}$ in the large gap region can be strongly dependent on the size and sign of $1/u$, while the curves essentially overlap in the small gap region around $\phi = 45$. This effect is related to the severity of the slope problem shown in Fig. 6, and can be traced back to the variation of the peak position in the self-energies displayed in Fig. 4 as a function of $1/u$.

Finally, we would like to illustrate another possible way to identify strong impurity scattering in ARPES experiments. Due to particle-hole mixing, the BCS spectral function Eq. (2) predicts that the ARPES intensity should not vanish abruptly when the normal state Fermi surface is crossed, but rather, a gradual decrease should occur, according to the dependence $v_{\mathbf{k}}^2 = (1 - \xi_{\mathbf{k}}/E_{\mathbf{k}})/2$ of the weight in the electron part of $A(\mathbf{k},\omega)$. At the same time, the peak in the spectral function approaches the Fermi energy only up to the minimum distance at $E_{\mathbf{k}_F} = -\Delta_{\mathbf{k}_F}$, and then disperses back to larger binding energy as $\mathbf{k}$ crosses the Fermi surface. Precisely this feature of BCS theory has recently been observed for the first time in Bi2212[19] making the case for the spectral function interpretation of the ARPES experiments on Bi2212 even stronger.

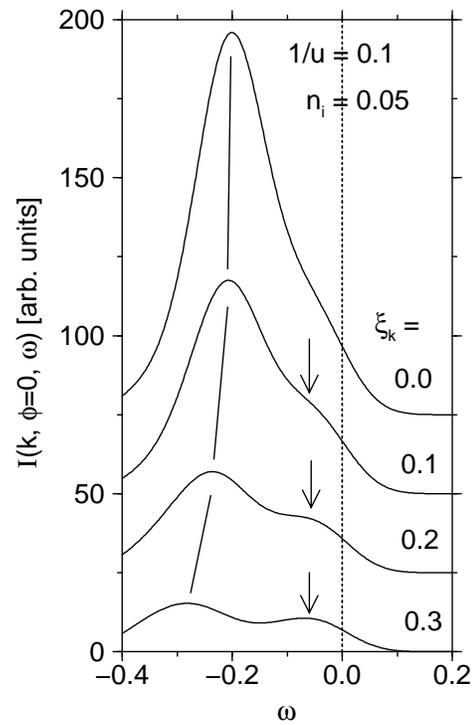

**Fig. 8.** Particle-hole mixing: Evolution of the EDC evaluated at $1/u = 0.1, n_i = 0.05$ (close to resonance) as the quasiparticle energy moves through the Fermi energy, $\xi_{\mathbf{k}} = 0.0, 0.1, 0.2, 0.3$. The lines indicate the dispersion of the electronic quasiparticle away from $\varepsilon_F$ as the Fermi energy is crossed. The arrows mark the incoherent, impurity induced peak near $\omega = 0$ which becomes more distinct as $\xi_{\mathbf{k}}$ grows.

Here, we demonstrate that impurity scattering may leave another unique fingerprint in the ARPES spectra when investigating the effect of the particle-hole mixing. The point is that as the Fermi surface is crossed, and the strength of the 'quasiparticle intensity', roughly given by $v_{\mathbf{k}}^2 = (1 - \xi_{\mathbf{k}}/E_{\mathbf{k}})/2$, weakens rapidly, the spectral weight introduced by the impurity self-energy in the vicinity of $\omega = 0$ decreases much slower. Hence, the incoherent part can show up as a second peak as soon as the coherent intensity (quasiparticle peak) is sufficiently small. This phenomena is demonstrated in Fig. 8 for a scattering strength $1/u = 0.1$ (near resonance, where the effect is strongest), and impurity concentration $n_i = 0.05$. One clearly observes how the quasiparticle peak disperses away from $\varepsilon_F$ as the Fermi surface is crossed, while the low-frequency tail present for $\xi_{\mathbf{k}} = 0$ progressively turns into a separate peak as $\xi_{\mathbf{k}}$ increases. We have plotted the EDCs without a normalization to show the drop in signal strength. Nevertheless, the intensity might still be enough to be observed.

## V. CONCLUSION

We have presented an analysis of ARPES data in the superconducting state of the high $T_c$ cuprates based on a simple phenomenological weak-coupling BCS model, which incorporates the normal state ARPES dispersion, and non-magnetic impurity scattering in the presence of a $d_{x^2-y^2}$ order parameter. We have shown, that it is extremely important to take into account the finite resolution of the analyzer in both momentum and frequency to obtain realistic results. In particular, the broadening by finite momentum resolution leads to a finite region of 'gaplessness' around the true node of the $d_{x^2-y^2}$ OP, if the shift of the leading edge of the spectrum, $\Delta_{sh}$, with respect to a non-superconducting reference sample (in electrical contact with the superconductor) is taken as the measure of the gap. This shift vanishes faster than linear in the Fermi surface angle $\phi$ around the $d_{x^2-y^2}$ node region, even though the true OP itself goes to zero linearly.

Non-magnetic impurities amplify the effect of the finite resolution. Additional spectral weight around $\omega = 0$ is introduced, leading to a further reduction of the shift $\Delta_{sh}$. At sufficiently large $n_i \approx 0.02 - 0.05$, this results in an extended region of apparent 'gaplessness' of $\delta\phi = \pm 7$ around the true node for a large range of moderate to strong impurity potential strengths. The zero frequency resonance responsible for a large residual DOS $N_{res}$ at $\epsilon_F$ manifests itself in the ARPES spectra in terms of a slope problem (particularly prominent in the large gap region around $[\pi,0]$), i.e., the slope of the ARPES spectra and that of the non-superconducting reference sample are far from parallel when the intensities are suitably normalized. There is a low frequency tail in the spectra reflecting the spectral weight from the impurity scattering near the resonance limit. We also demonstrated that this spectral weight could be identified by looking at the effect of particle-hole mixing for **k** values which are outside the normal state Fermi surface. In this case, the incoherent spectral weight around $\omega = 0$ leads to a second peak in the ARPES spectra with a weight comparable to that of the quasiparticle part.

One point about the impurity related parameters used in our calculations has to be born in mind: for $n_i \geq 0.02$, $|1/u| \leq 0.8$, our model also leads to a significant reduction in $T_c$ (see Ref. 4) up to 50% at $n_i = 0.05$. Hence, if an extended region of nodes is observed in an ARPES experiment, this should be accompanied by a substantial $T_c$ reduction compared to a pure sample where such a 'nodeless' region is absent. If this is not the case, it might well be that the sample surfaces have a higher impurity concentration than the bulk, so that the $T_c$ suppression is determined by a different $n_i$ than the one that affects the ARPES spectra. Note also, that NMR experiments on Bi2212 crystals grown by the same group as the ones used in the ARPES experiments Ref. 3, have indicated the presence of a large $N_{res}$ without explicitly doped impurities.[20] This result is somewhat puzzling, since the reported value of $N_{res}$ (24% of the normal DOS) would imply a $T_c$ reduction of at least 20% (according to our model calculations), but the $T_c$ of the sample was still 86K, which is within 10% of the best Bi2212 values in the literature. Nevertheless, if this residual DOS is due to non-magnetic impurity scattering, one would expect the presence of an extended nodeless region in ARPES measurements on those samples.

A final word about resolution effects: Improved frequency resolution would be beneficial for the understanding of the ARPES line shape around $(\pi,0)$, since due to the proximity of the saddle point, momentum broadening is very weak in this region, and one might be able to observe more features of the true spectral function. On the other hand, better momentum resolution would definitely help to improve the accuracy in determining the size of the gap since in the most interesting region around the $d_{x^2-y^2}$ node, momentum resolution broadening is clearly the limiting factor due the rapid quasiparticle dispersion.

The author would like to thank M. R. Norman for many useful comments and discussions, and acknowledges financial support by the NSF (DMR-91-20000) through the Science and Technology Center for Superconductivity.

[1] B. G. Levi, Phys. Today **49**, 19 (1996).
[2] Z.-X. Shen et al., Phys. Rev. Lett. **70**, 1553 (1993).
[3] H. Ding et al., Phys. Rev. Lett. **74**, 2784 (1995); H. Ding et al., Phys. Rev. Lett. **75**, 1425 (1995).
[4] R. Fehrenbacher, preprint cond-mat/9512105.
[5] L. P. Gorkov, Pis'ma Zh. Eksp. Teor. Fiz. **40**, 351 (1984), [Sov. Phys. JETP Lett. **40**, 1155 (1985)]; K. Ueda and T. M. Rice, in *Theory of Heavy Fermions and Valence Fluctuations*, edited by T. Kasuya and T. Saso, page 267, Berlin, 1985, Springer.
[6] C. J. Pethick and D. Pines, Phys. Rev. Lett. **57**, 118 (1986); S. Schmitt-Rink, K. Myake, and C. M. Varma, Phys. Rev. Lett. **57**, 2575 (1986).
[7] P. J. Hirschfeld, P. Wölfle, and D. Einzel, Phys. Rev. B **37**, 83 (1988); H. Monien, K. Scharnberg, and D. Walker, Solid State Commun. **63**, 263 (1987).
[8] R. Fehrenbacher and M. R. Norman, Phys. Rev. B **50**, 3495 (1994).
[9] M. Randeria et al., Phys. Rev. Lett. **74**, 4951 (1995).
[10] D. S. Dessau et al., Phys. Rev. Lett. **71**, 2781 (1993).
[11] R. Fehrenbacher and M. R. Norman, Phys. Rev. Lett. **74**, 3884 (1995).
[12] M. R. Norman et al., Phys. Rev. B **52**, 615 (1995).
[13] K. Gofron et al., J. Phys. Chem. Solids **54**, 1193 (1993).
[14] T. Xiang and J. M. Wheatley, Phys. Rev. B **51**, 11721 (1995).
[15] D. S. Dessau et al., Phys. Rev. Lett. **66**, 2160 (1991).
[16] H. Ding et al., to be published in Spectroscopic Studies of Superconductors, edited by I. Bozovic and D. van der Marel (SPIE, Bellingham, 1996).
[17] J. Giapintzakis, Phys. Rev. B **50**, 15967 (1994).
[18] J. C. Campuzano, private communication.
[19] J. C. Campuzano et al., preprint cond-mat/9602119.
[20] K. Ishida et al., J. Phys. Soc. Jpn. **63**, 1104 (1994).